\documentclass[prl,twocolumn,superscriptaddress]{revtex4}
\usepackage[ansinew]{inputenc}
\usepackage[dvips]{graphicx}
\usepackage{amsmath}
\usepackage{amsthm}
\usepackage{layout}
\usepackage{float}
\usepackage{subfigure}
\usepackage{amsfonts}
\usepackage{amssymb}
\usepackage{txfonts}

\newcommand{\ket}[1]{\left\vert#1\right\rangle}

\begin{document}
\vskip4pc

\clearpage

\title{A high-speed tunable beam splitter for feed-forward photonic quantum information processing}

\author{Xiao-song Ma}
\affiliation{Institute for Quantum Optics and Quantum Information (IQOQI), Austrian
Academy of Sciences, Boltzmanngasse 3, A-1090 Vienna, Austria}
\affiliation{Vienna Center for Quantum Science and Technology, Faculty of Physics, University of Vienna, Boltzmanngasse 5, A-1090 Vienna,
Austria}\email{xiaosong.ma@univie.ac.at}
\author{Stefan Zotter}
\affiliation{Institute for Quantum Optics and Quantum Information (IQOQI), Austrian
Academy of Sciences, Boltzmanngasse 3, A-1090 Vienna, Austria}
\affiliation{Vienna Center for Quantum Science and Technology, Faculty of Physics, University of Vienna, Boltzmanngasse 5, A-1090 Vienna,
Austria}
\author{Nuray Tetik}
\affiliation{Institute for Quantum Optics and Quantum Information (IQOQI), Austrian
Academy of Sciences, Boltzmanngasse 3, A-1090 Vienna, Austria}
\affiliation{Vienna Center for Quantum Science and Technology, Faculty of Physics, University of Vienna, Boltzmanngasse 5, A-1090 Vienna,
Austria}
\author{Angie Qarry}
\affiliation{Institute for Quantum Optics and Quantum Information (IQOQI), Austrian
Academy of Sciences, Boltzmanngasse 3, A-1090 Vienna, Austria}
\affiliation{Vienna Center for Quantum Science and Technology, Faculty of Physics, University of Vienna, Boltzmanngasse 5, A-1090 Vienna,
Austria}
\author{Thomas Jennewein}
\affiliation{Institute for Quantum Optics and Quantum Information (IQOQI), Austrian
Academy of Sciences, Boltzmanngasse 3, A-1090 Vienna, Austria}
\affiliation{Institute for Quantum Computing and Department of Physics and Astronomy, University of Waterloo, 200 University Avenue West, Waterloo, ON, N2L 3G1, Canada}
\author{Anton Zeilinger}
\affiliation{Institute for Quantum Optics and Quantum Information (IQOQI), Austrian
Academy of Sciences, Boltzmanngasse 3, A-1090 Vienna, Austria}
\affiliation{Vienna Center for Quantum Science and Technology, Faculty of Physics, University of Vienna, Boltzmanngasse 5, A-1090 Vienna, Austria}
\email{anton.zeilinger@univie.ac.at}

\begin{abstract}
We realize quantum gates for path qubits with a high-speed, polarization-independent and tunable beam splitter. Two electro-optical modulators act in a Mach-Zehnder interferometer as high-speed phase shifters and rapidly tune its splitting ratio. We test its performance with heralded single photons, observing a polarization-independent interference contrast above $95\%$. The switching time is about 5.6 ns, and a maximal repetition rate is 2.5 MHz. We demonstrate tunable feed-forward operations of a single-qubit gate of path-encoded qubits and a two-qubit gate via measurement-induced interaction between two photons.
\end{abstract}


\maketitle
\section{Introduction}
In the rapidly growing area of photonic quantum information processing~\cite{Gisin2002, Kok2007, Scarani2009, Ladd2010}, beam splitter is one of the basic building blocks~\cite{Reck1994}. It is crucial for building single-photon, as well as two-photon gates. The measurement-induced nonlinearity realized by two-photon interference~\cite{Hong1987} on a beam splitter enables the interaction between photons and hence along with single photon gates allows the realization of quantum dense coding~\cite{Mattle1996}, quantum teleportation~\cite{Bouwmeester1997}, entanglement swapping~\cite{Pan1998,Jennewein2002}, multi-photon interferometers~\cite{Pan2008} and many two-photon quantum gates~\cite{Kok2007}.

Particularly quantum computation~\cite{Knill2001,Raussendorf2001,Prevedel2007}, quantum metrology~\cite{Higgins2007} and quantum simulation~\cite{Lanyon2010, Ma2011} require high-speed, tunable quantum gates (beam splitters) controlled by the signals from feed-forward detections to enhance their efficiency and speed. Also these special beam splitters can be used in a single-photon multiplexing scheme to improve the quality of single photons~\cite{Migdall2002, Shapiro2007,Ma2011b}. Since the polarization of a photon is a convenient and widely-used degree of freedom for encoding quantum information, such a beam splitter is preferably polarization-independent and conserves coherence. Therefore, a high-speed, polarization-independent, tunable beam splitter for photonic quantum information processing is highly desirable. In classical photonics, although fast optical modulators based on electro-optical materials embedded in Mach-Zehnder interferometers (MZI) are already at an advanced stage~\cite{Eldada2001}, the polarization preference of these devices makes them unsuitable for quantum information processing. Recently, several photon switches have been demonstrated~\cite{Spagnolo2008, Ng2008, Hall2010}. However, the demonstrations of feed-forward operations on path-encoded qubits still remained a challenge.

\begin{figure}[h!]
\centerline{\includegraphics[width=0.5\textwidth]{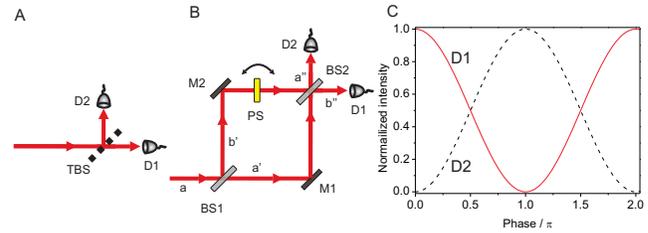}}\caption{(Color online.) The concept of a tunable beam splitter. \textbf{A}. The splitter ratio (the ratio between the transmissivity and reflectivity) of a tunable beam splitter (TBS) can be adjusted. Therefore, the counts of detectors D1 and D2 vary according to the splitting ratio. \textbf{B}. The realization of a TBS with a Mach-Zehnder interferometer, which consists of two beam splitters (BS1 and BS2), two mirrors (M1 and M2) and a phase shifter (PS). The splitting ratio can be tuned by adjusting PS. \textbf{C}. The normalized intensities of D1 (red solid curve) and D2 (black dashed curve) are plotted as a function of the phase.} \label{TBSscheme}
\end{figure}

The operational principle of our high-speed, tunable beam splitter (TBS) is the following. In order to build a tunable beam splitter (Fig.\ref{TBSscheme}A), one needs a knob to adjust its splitting ratio. One of the possibilities is to vary the phase of a Mach-Zehnder interferometer (MZI), as shown in Fig.\ \ref{TBSscheme}B. The splitting ratio can be tuned by adjusting the phase of interferometer. This is represented as the intensity modulations of the outgoing beams (Fig.\ \ref{TBSscheme}C). For instance, if the phase is $0$ there is no beam splitting, and if the phase is $\frac{\pi}{2}$ the splitting ratio is 1.

\section{Experiment}
\begin{figure}[h!]
\centerline{\includegraphics[width=0.45\textwidth]{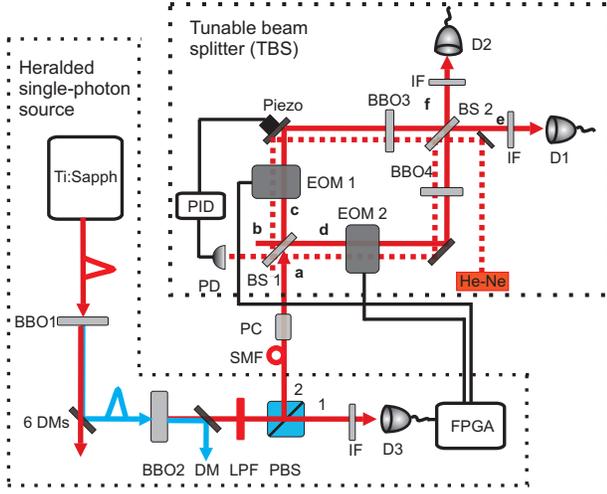}}\caption{(Color online.) Experimental
setup for performing feed-forward operation of a path-encoding single-qubit gate with heralded single photons. Femtosecond laser pulses from a Ti:Sapphire oscillator ($\lambda$ = 808 nm) are up-converted with a $\beta$-barium borate crystal (BBO1). The up-converted pulses and the remaining fundamental pulses are separated with six dichroic mirrors (6 DMs). A correlated photon pair is generated from BBO2 via spontaneous parametric down-conversion. By detecting photon 1 in the transmitting arm of the polarizing beam splitter (PBS) with an avalanche photon detector (D3), we herald the presence of photon 2 in the reflecting arm of the PBS. Photon 2 is delayed with an optical fiber and the polarization rotation of the fiber is compensated by a polarization controller (PC). Then the photon is sent to the tunable beam splitter (TBS). Both photons are filtered by using interference filters (IF) with 3 nm bandwidth centered around 808 nm. The output signal of the detection of photon 1 is used to trigger two EOMs. The time delay between trigger pulse and the arrival of photon 2 at the EOMs is adjusted via the field-programmable gate array (FPGA) logic, which allows the feed-forward operations. Photon 2 is detected by D1 or D2 at the output of the TBS. Note that the interferometer is actively stabilized by using an auxiliary He-Ne laser (He-Ne), a silicon photon detector (PD), an analogue proportional-integral-derivative (PID) control circuit and a ring-piezo transducer.} \label{TBS1}
\end{figure}

Here we present an experimental realization of such a high-speed, polarization-independent, tunable beam splitter based on bulk electro-optic modulators (EOM) embedded in a Mach-Zehnder interferometers, as shown in Fig.\ \ref{TBS1}. The TBS consists of two 50:50 beam
splitters, mirrors and most importantly two EOMs (one in each arm of
the interferometer). EOMs vary the phase of transmitted light by the
application of an electric field inducing a birefringence in a
crystal. Here we use Rubidium Titanyl Phosphate (RTP) crystals as
the electro-optic material for our EOMs. Since it lacks
piezo-electric resonances for frequencies up to 200 kHz~\cite{Leysop2009}
and shows very rare resonances up to 2.5 MHz, it is suitable to drive the EOMs at high repetition rate.

The interferometer is built in an enclosed box made from
acoustic isolation materials in order to stabilize the phase
passively. Additionally, an active phase stabilization system is
implemented by using an auxiliary beam from a power-stabilized He-Ne laser counter-propagating through the whole MZI, as shown by the red dashed line in Fig.\ \ref{TBS1}. This beam has a small transversal displacement from the signal beam, and picks up any phase fluctuation in the interferometer. The corresponding intensity variations of the output He-Ne laser beam are measured with a silicon photon detector (PD) and the signal is fed into an analogue proportional-integral-derivative (PID) control circuit. A ring-piezo transducer attached to one of the mirrors in the MZI is controlled by this PID and compensates the phase fluctuations actively. Since the wavelength of the He-Ne laser (about 633~nm) is smaller than the wavelength used in the quantum experiment, it implies higher sensitivity to fluctuations of the phase.

The optical axes of both RTP crystals are oriented along $45^{\circ}$ and the voltages applied to them are always of the same amplitudes
but with opposite polarities. The scheme of using the two EOMs is crucial, because the tunability of this high-speed tunable beam splitter relies on first-order interference. By employing EOMs in both arms, the polarization states of photon 2 in path \textbf{c} and in path \textbf{d} remain indistinguishable and hence allow first-order interference. The voltages applied to the EOMs tune the relative phase of the MZI. The phase of the MZI is defined to be $0$ when all the photons entering from input path \textbf{a} exit into the output path \textbf{f}. This also corresponds to the phase locking point of the He-Ne beam.

The functionality of the TBS is best seen by describing how the quantum
states of polarization and path of the input photon evolve in the
MZI.  Since the optical axes of the EOMs are along $+45^{\circ}$/$-45^{\circ}$, we decompose the input
polarization into the $\ket{+}$/$\ket{-}$ basis, with $\ket{+}$ ($\ket{-}$) being the eigenstate to the $+45^{\circ}$ ($-45^{\circ}$) direction. The arbitrarily polarized input state in spatial mode \textbf{a} is $\ket{\Psi}
=(\alpha\ket{+}+\beta\ket{-})\ket{a}$, where $\alpha$ and $\beta$ are complex amplitudes and normalized ($|\alpha|^{2}+|\beta|^{2}=1$). It evolves as:
\begin{eqnarray}
\ket{\Psi} \;     &\underrightarrow{\textrm{TBS}} &
                   \sin{\frac{\phi(U)}{2}} e^{\mathrm{i}(\frac{3\pi}{2}-\frac{\phi(U)}{2})}(\alpha\ket{+}-\beta\ket{-})\ket{e}\; \nonumber \\
                   & & +\cos{\frac{\phi(U)}{2}} e^{\mathrm{i}(\frac{\pi}{2}+\frac{\phi(U)}{2})}(\alpha\ket{+}+\beta\ket{-})\ket{f},\label{TBSeq1}
\end{eqnarray}
where $\phi(U)$ is the voltage dependent birefringence phase given
by the EOMs.  From Eq.\ \ref{TBSeq1}, it is straitforward to see that one can
tune the splitting ratio by
varying $\phi(U)$.  The transmissivity ($T$) and reflectivity ($R$) are defined as the probabilities of detecting photons, entering from mode \textbf{a}, in the outputs of spatial modes \textbf{f} and \textbf{e}: $T=\cos^2{\frac{\phi(U)}{2}}$ and $R=\sin^2{\frac{\phi(U)}{2}}$. Additionally, the input polarization will be rotated from
$\alpha\ket{+}+\beta\ket{-}$ to $\alpha\ket{+}-\beta\ket{-}$ in the
spatial mode \textbf{e}. This polarization rotation can be dynamically
compensated with an additional EOM on path \textbf{e} applied with a half
wave voltage.

We test the performance of our TBS with a heralded single-photon source sketched in
Fig.\ \ref{TBS1}.  By using the correlation of the emission time of the photon
pair generated via spontaneous parametric down-conversion (SPDC), we herald the presence of one photon with
the detection of its twin, the trigger photon. In order to employ the feed-forward technique, the detection of the
trigger photon is used to control the EOMs in the TBS, which
operates on the heralded single photon.
\begin{figure}[ht]
\centerline{\includegraphics[width=0.45\textwidth]{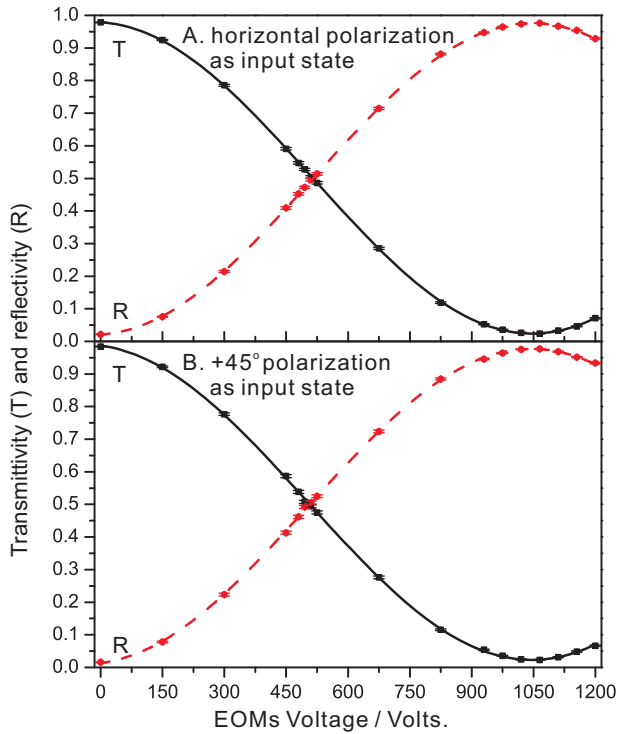}}
\caption{(Color online.) Demonstration of the tunable and polarization-independent feed-forward operation of a single-qubit gate of path-encoded qubits. Experimental results for \textbf{A} horizontally and \textbf{B} $+45^{\circ}$ polarized input single photons. The black squares (red circles) are the
values of the transmissivity (reflectivity) of the ultrafast tunable beam
splitter, which are calculated from the coincidence counts between
D1 and D3 (D2 and D3) and fitted with a black solid (red dash) sinusoidal curve. Error bars represent statistical errors of $\pm$1 standard deviation and are comparable with the size of the data points.} \label{TBS2}
\end{figure}

Femtosecond laser pulses from a Ti:Sapphire oscillator are up-converted with a 0.7 mm $\beta$-barium borate crystal (BBO1) cut for
type-I phase-matching. This produces vertically polarized
second harmonic pulses. The up-converted pulses and the remaining fundamental pulses are separated with six dichroic mirrors (6 DMs). The
collinear photon pairs are generated via SPDC from a 2 mm BBO
crystal (BBO2) cut for collinear type-II phase-matching. The UV pulses are removed from
the down converted photons with a dichroic mirror (DM) and long pass
filter (LPF). The photon pair is separated by a polarizing beam
splitter (PBS) and each photon is coupled into a single-mode fiber
(SMF).

The transmitted photon, photon 1, is detected by an avalanche photon detector (D3), which serves as the trigger to control the EOMs.
The reflected photon, photon 2, is delayed in a 100~m single-mode fiber and then sent through the TBS.  The on-time of the EOMs is 20 ns, which requires fine time delay adjustment
of the detection pulse from D3 used to trigger the EOMs. It is achieved with a field-programmable gate array (FPGA). To use an on-time of 20 ns is a compromise of the performance of EOMs and the repetition rate of our pulsed laser system. On one hand, the rising and falling times of the EOM are about 5 ns (see below) and hence the on-time of the EOMs has to be longer than 10 ns. Experimentally, we find that it has to be at least 20 ns for achieving a polarization switching contrast above 98\%. On the other hand, the repetition rate of our laser is about 80 MHz and it corresponds to a period of 12.5 ns. To avoid switching the uncorrelated photons generated from distinct laser pulses, the on-time should be comparable to 12.5 ns. In our experiment, we use a relative low pump power for SPDC and the chance of generating two pairs of photons in consecutive pulses is extremely low. Therefore, 20 ns on-time is the suitable choice for our application.

In order to maximize the operation quality of the TBS, the path length difference between the two arms of the MZI has be minimized. In addition, two pairs of crossed
oriented BBO crystals (BBO3 and BBO4) are placed in each arm of the
MZI in order to compensate the unwanted birefringence induced by
optical elements. Photon 2 is detected by avalanche photon detectors D1 or D2 at the output of the TBS.

\begin{figure}[ht]
\includegraphics[width=0.4\textwidth]{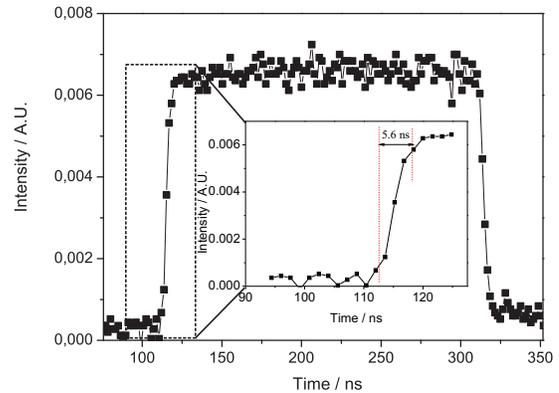}
\caption{Measured switching response of the tunable beam splitter. This is measured with a continuous wave laser and a Si photon
detector. A rise time of 5.6 ns is observed, referring to the time for the signal to rise from 10\% to 90\%. The switching starts at an offset time of 110.4 ns. Note that, rise and fall time are typical parameters used to quantify the response speed of devices and are defined as the time required for a signal to change from 10\% and 90\% of the step height for the rise time (90\% and 10\% for the fall time).} \label{TBS3}
\end{figure}

One important feature of our TBS is polarization independence, i.e.\ it works in a similar way for all polarizations. This is tested
by preparing the polarization state of photon 2 with a fiber
polarization controller (PC), as shown in Fig.\ \ref{TBS2}A for horizontal polarization and Fig.\ \ref{TBS2}B for $+45^{\circ}$ polarization. We fit sinusoidal curves
to the measurement results and determine that the visibilities for horizontal and $+45^{\circ}$ are $95.9\%\pm0.2\%$ and $95.3\%\pm0.3\%$, respectively. It is defined as $V=(\textrm{R}_\textrm{Max}-\textrm{R}_\textrm{Min})/(\textrm{R}_\textrm{Max}+\textrm{R}_\textrm{Min})$, where $\textrm{R}_\textrm{Max}$ and $\textrm{R}_\textrm{Min}$ are the maximum and minimum of the reflectivity. For input path \textbf{b}, we
have observed the same results, which confirms the usefulness of this device
to perform two-photon interference experiments. The polarization fidelity of the switchable beam splitter for an arbitrary state is above 98\%.

Two further important requirements of a TBS are high-frequency operation
and short rise and fall times. These are challenging to achieve with today's electro-optical active crystals and the driving electronics. We use the particular advantages of RTP crystals and drive the EOMs with
frequencies up to 2.5 MHz.  We measure the rise/fall time  (10\% to 90\% of signal amplitude transition time) of the TBS with a continuous wave laser with a Si photon detector
and an oscilloscope. As shown in Fig.\ \ref{TBS3}, the rise time of a $\pi$-phase modulation is about 5.6 ns. Note that the fall time is about the same.

For an optical two-qubit entangling gate, Hong-Ou-Mandel (HOM) two-photon interference~\cite{Hong1987} is at the heart of many quantum information processing protocols, especially for photonic quantum computation experiments (C-phase gate~\cite{Langford2005, Kiesel2005, Okamoto2005}, entanglement swapping~\cite{Pan1998, Jennewein2002}, etc.).  In order to use our optical circuitry for more complicated quantum logical operations, it is crucial to demonstrate with it such two-photon interference phenomena. Therefore, we also carried out a tunable two-photon interference experiment with our TBS.

We use a pair of photons generated from one source and send them to the TBS. We vary the path length difference between these two photons with a motorized translation stage mounted on one of the fiber coupling stages and measure the two-fold coincidences between two detectors placed directly behind two outputs of the TBS (\textbf{e} and \textbf{f} in Fig.\ \ref{TBS1}). The phase of the MZI and hence the reflectivity (and transmissivity) of the TBS is varied by applying different voltages on the EOMs. In case of a $\pi/2$ phase, the TBS becomes a balanced beam splitter and the distinguishability of two input photons' spatial modes is erased. In consequence, the minimum of the coincidence counts occurs for the optimal temporal overlap (with the help of suitable individual fiber delays) of the two photons, and HOM two-photon interference with a visibility of $88.7\%\pm3.8\%$ has been observed. In case of a $0$ phase, the whole TBS represents a highly transmissive beam splitter and the two photons remain distinguishable in their spatial modes. Correspondingly, one obtains path length difference insensitive coincidence counts. This is in agreement with complementarity, where in principle no HOM interference can be observed. In ref~\cite{Ma2011b}, some results are presented in different context.

The insertion loss of the TBS is about 70\%, which is mainly due to single-mode fiber coupling and the Fresnel loss at optical surfaces. With current technology, it is in principle possible to improve the transmission of a TBS to about 95\%, including Fresnel loss on each optical surface (0.5\% per surface) and a finite fiber coupling efficiency of 98\%.

\section{Conclusion}
In conclusion, we experimentally demonstrate tunable feed-forward operations of a single-qubit gate of path-encoded qubits and a two-qubit gate via measurement-induced interaction between two photons by using a high-speed polarization-independent tunable beam splitter. We have shown its unique advantages, such as excellent fidelity, high speed, and high repetition rate, operating for photons at about 800 nm. The active stabilization allows continuous usage of this tunable beam splitter over periods of many days. This systems is well suited for experimental realizations of quantum information processing and fundamental quantum physics tests. A future challenge will be to utilize state-of-art micro-optics technology, such as integrated photonic quantum circuits on a silicon chip \cite{Politi2008, Matthews2009}, for developing compact and scalable quantum technologies based on photons.

\section*{Acknowledgements}
 We thank J. Kofler, P. Walther and W. Naylor for discussions and for improving the manuscript. We acknowledge support from the European Commission, Q-ESSENCE (No. 248095), ERC Senior Grant (QIT4QAD), JTF, SFB FoQus and the Doctoral Program CoQuS of the Austrian Science Foundation (FWF).

\end{document}